\documentclass[aps,prb,twocolumn,superscriptaddress]{revtex4-2}
\usepackage{graphicx}
\usepackage{picture}
\usepackage{xcolor}
\usepackage{color}
\usepackage{tikz}
\usepackage{amsmath}
\usepackage{rotating}
\usepackage{comment}

\begin{document}
\title{Dynamics away from Equilibrium of a Disordered Binary Alloy} 

\author{Aly Abuelmaged}
\affiliation{Department of Physics, University at Buffalo SUNY, Buffalo, New York 14260, USA}
\author{Eric Dohner}
\affiliation{Earth Resources Technology Inc, Greenbelt, MD, 20770, USA}
\author{Shang-Jie Liou}
\affiliation{Department of Physics, University at Buffalo SUNY, Buffalo, New York 14260, USA}
\author{Herbert F Fotso}
\affiliation{Department of Physics, University at Buffalo SUNY, Buffalo, New York 14260, USA}

\begin{abstract}

We study the nonequilibrium dynamics of a binary disordered alloy when it is subjected to an interaction quench. Our study uses a nonequilibrium embedding scheme (DMFT+CPA) that combines the capacity of DMFT (dynamical mean field theory) to treat strongly correlated systems and the capacity of CPA (coherent potential approximation) to treat disordered systems, to effectively address the interplay of disorder and interaction for the nonequilibrium system. Our solution is applied on the equilibrium binary disordered and interacting alloy for the calculation of the density of states. This equilibrium density of states shows the modifications of the disorder-driven gap by the interaction and that of the interaction-driven Mott gap by the disorder. Next, we assess the relaxation of the system that is initially in equilibrium at a given temperature, and then abruptly has its interaction strength changed from zero to a finite value at which it is kept constant. The system undergoes a short time transient and then settles into a long-time state in a process that depends in a nontrivial manner on the disorder and interaction strength. In addition, we calculate the effective temperature of the system after it has gone through its initial transient and settled into its long-time state. The long-time effective temperature is increased with increasing final interaction strength. However, for a given final interaction, the final temperature can be tuned by the disorder strength over a broad range, with stronger disorder leading to lower temperature. Altogether, the results illustrate the nontrivial interplay between disorder and interaction for a binary alloy away from equilibrium. 
\end{abstract}

\maketitle

\section{Introduction}
\noindent Disordered many-particle systems remain an area of great interest due to the intriguing properties that arise from the non-trivial effects arising from the presence of disorder. For example, disorder is known to give rise to Anderson localization that has now been demonstrated in a broad variety of wave systems~\cite{AndersonPhysRev1958, LeeRamakrishnanRevModPhys1985, ElihuBook_50years_of_AL_2010, MottAdvPhys1967, THOULESS1974, KirkpatrickPRB1985, LagendijkPhysToday2009, SajeevPRL1984, Wiersma1997, Segev2013, 1b, 2b, 3c, 4b}. In addition, the interplay of disorder and interaction has been the subject of great scrutiny~\cite{PalHuse_prb2010, BASKO20061126, AbaninAltmanBloch_RMP2019, OganesyanHuse_PRB2007, NandkishoreHuseAnRevCondMattPhys2015}. Despite all this interest and despite their expected importance, the nonequilibrium dynamics of systems that feature both strong electron-electron interaction and disorder remains generally understudied. This is in part due to the combined complexities of correlated interacting systems and that of disordered systems away from equilibrium.  
On the one hand, the dynamical mean field theory (DMFT) has been established as an effective embedding scheme to address strongly correlated systems~\cite{33a, 34a, 35a, 36a, 37a}. On the other hand, the coherent potential approximation (CPA) is a widely used embedding scheme to treat disordered many-particle systems~\cite{38a, 39a, 40a, 41a}. Both methods have been previously combined to treat strong correlation and disorder on equal footing in equilibrium, with the resulting algorithms producing numerous valuable results~\cite{UlmkeJanisVollhardt, JanisVollhardt_PRB1992, LombardoEtAl2006, LaadMHartmannPRN2001, PRB_DMFT_AndersonHubbard_WehVollhardtEtAl2021}.  In addition, both CPA and DMFT have now been separately adapted to the treatment of the nonequilibrium dynamics of correlated systems~\cite{48a, 49a, 50a, 54a, 55a, 56a, 68a, 57a} and of disordered systems respectively~\cite{66a, 67a}.
Recently, the nonequilibrium DMFT+CPA, has been introduced as an embedding scheme combining both DMFT and CPA for the treatment of the nonequilibrium dynamics of disordered correlated systems~\cite{DMFT_CPA_NonEq2022, dohner2023thermalization}.
The method and some of its variations have since been applied to various situations including quantum transport~\cite{DMFTCPA_PWerner} and strong-field driven systems~\cite{DMFTCPA_Arrigoni}. In typical studies, systems incorporating both site-disorder and interaction are described by the Anderson-Hubbard model and its extensions. Disorder is then represented by a site dependent potential that takes on random values following a given probability distribution function defining the type of disorder. For a disordered binary alloy, when disorder is characterized by a binary distribution, the coherent potential approximation effectively captures the onset of a disorder-induced metal-to-insulator transition while the interaction-driven formation of the Mott gap is captured by DMFT. Although the interplay of these two transitions has been investigated in equilibrium~\cite{LaadMHartmannPRN2001, LombardoEtAl2006, PRB_DMFT_AndersonHubbard_WehVollhardtEtAl2021}, its impact on the nonequilibrium dynamics has received little attention to date. 

In the present paper, we consider the nonequilibrium dynamics of a disordered correlated binary alloy, described by the Anderson-Hubbard model with binary disorder, under an interaction quench whereby the interaction is abruptly switched from one initial value to another. Our nonequilibrium solution is first used to examine the equilibrium density of states of the system and its dependence on both the interaction and the disorder strength. We find that disorder delays the opening of the Mott gap to stronger interactions in agreement with previous results, including the recently described effects of interactions on disordered metals~\cite{ArinnaCiuchi2025} Vice-versa, the interaction delays the opening of the disorder gap to stronger disorder. Across the interaction quench, we find that disorder plays a non-trivial role on the dynamics and on the long-time state.
For weak interaction, we observe that the final kinetic energy is increased with the disorder strength. As the interaction increases, we observe at moderate interactions, a reversal in the initial trend for the kinetic energy, with increased disorder strength leading to lower kinetic energy. 
Altogether, the results illustrate the nontrivial interplay between disorder and interaction for a binary alloy.

The rest of the paper is organized as follows. In section ~\ref{sec:ModelMethod}, we present our model and discuss the method for solving the interacting disordered system out of equilibrium. In section ~\ref{sec:results}, we present our results, first for the equilibrium density of states. These are obtained in the weak-to-moderate coupling and disorder regimes, obtained with the time domain formalism that bypasses any form of analytical continuation. The discussion of the equilibrium density of states is followed by the analysis of the dynamics of the system after the interaction quench and the calculation of the long-time temperature of the system as a function of the interaction and the disorder strength. We end with our conclusion in section ~\ref{sec:conclusion}.

\section{Model and Method}
\label{sec:ModelMethod}

\subsection{Model}

\noindent We consider  a disordered correlated system that is described in equilibrium by the single-band Anderson-Hubbard model, the Hamiltonian of which can be written as:
    \begin{eqnarray}
    \label{eq:AndersonHubbard}
         H  = - \sum_{\langle i j \rangle \sigma} &t_{ij}& (c^\dagger_{i \sigma} c_{j \sigma} + h.c.) + \sum_{i} U n_{i \uparrow} n_{i \downarrow} \nonumber \\
        & + &  \sum_{i \sigma}\left(V_i - \mu \right) n_{i \sigma}.
    \end{eqnarray}
It describes electrons on a lattice where they are allowed to hop between nearest neighboring sites, interact strongly when they doubly occupy a site, and have a random disorder potential $V_i$  on site $i$. $\mu$ is the chemical potential. $t_{ij} = t_{hop}$ is the hopping amplitude between nearest neighboring sites denoted by $\langle i j \rangle$. We work in units where $c = \hbar = e = 1.$
$c^\dagger_{i \sigma}$ ($c_{i \sigma}$) is the creation (annihilation) operator for a particle of spin $\sigma =\uparrow,\downarrow$ at site $i$. $U$ is the Coulomb interaction strength at a doubly occupied site. $n_{i, \sigma}= c^\dagger_{i \sigma}c_{i \sigma}$ is the number operator for particles of spin $\sigma$ at site $i$.

$V_i$ is randomly distributed according to a probability distribution $P(V_i)$. Here we are interested in binary disorder whereby the disorder potential at site $i$ can randomly take one of two values: $W/2$ or $-W/2$. $W$ defines our disorder strength and $P(V) = \frac{1}{2}\delta(V-W/2) + \frac{1}{2}\delta(V+W/2)$.
We will employ the shorthand notation $<A(t,t')>_{\{V\}}=\int dV_i P(V_i) A(t,t')$ to denote disorder averaging of the quantity $A(t,t')$. Our calculations will address the Anderson-Hubbard model on the Bethe lattice with a large coordination number $z  \rightarrow \infty$ and the hopping is renormalized such that $t_{hop} \to t^*/\sqrt{z}$. We use $t^* = 0.25$ to set our energy and time units, and our system is kept at half-filling. 

For the nonequilibrium problem, with the disorder strength $W$ kept at a fixed value, at time $t_{quench}$, the interaction strength is abruptly switched from an initial value of $U_1 = 0$ to a final value $U_2 = U$ at which it is kept constant. 

\subsection{Nonequilibrium DMFT+CPA embedding scheme for interaction and disorder}

\noindent The nonequilibrium solution for the many-body problem is obtained using the Kadanoff-Baym-Keldysh formalism,  generally depicted graphically by an $L$-shaped contour~\cite{62a, 63a, 64a, 65a}, whereby the system is evolved forward from an initial time $t_{min}$ up to a maximum time $t_{max}$, backward to $t_{min}$, and then vertically along the imaginary time axis extending from $t_{min}$ to $t_{min} - i\beta$ where $\beta$ is the initial temperature at which the system is initially in equilibrium at time $t_{min}$.

The fundamental quantity of the formalism is the contour ordered Green's function between sites $i$ and $j$ for a particle of spin $\sigma$:
\begin{equation}
    G^c_{i, j, \sigma}(t,t') = -i\langle \mathcal{T}_c c_{i \sigma}(t) c^{\dagger}_{j \sigma}(t') \rangle .
    \label{eq:GContour} 
\end{equation}
The $t$ and $t'$ time variables are taken on any point of the contour. $\mathcal{T}_c$ represents time ordering along the Keldysh contour. 
The contour-ordered Green's function is also related to the lesser ($G^<$) and greater ($G^>$) Green's functions by:
\begin{equation}
        G^c_{i, j, \sigma}(t, t')  =  \theta_c(t, t')G^>_{i,j, \sigma}(t, t') + \theta_c(t', t)G^<_{i,j, \sigma}(t, t'). \;\;\;\;\label{eq:GContour2} 
\end{equation}
Here, $\theta_c(t,t')$ defines the contour-ordered Heaviside function. It orders time with respect to the contour: $\theta_c(t,t') = 1$ if $t$ is ahead of $t'$ on the contour, and $\theta_c(t,t') = 0$ otherwise. Hereafter we drop the superscript $c$ on the contour-ordered Green's function. Thus, any correlation function not identified as a particular type (e.g., not identified as $G^<$) should be understood to refer to the full contour-ordered Green's function.

From the contour-ordered Green's function, we can extract the other Green's functions such as the lesser, the greater, or the retarded ($G^R$) Green's functions. With the retarded Green's function defined by:
 \begin{equation}
 G^R_{i,j, \sigma}(t,t')  =  -i\theta(t-t')\langle \{c_{i\sigma}^{\phantom\dagger}(t), c^{\dagger}_{j\sigma}(t')\}\rangle \label{eq:GRetarded}.
\end{equation}

The nonequilibrium DMFT+CPA approach maps the lattice problem onto an impurity embedded into a self-consistently determined bath. Details of the algorithm were discussed extensively in earlier work~\cite{DMFT_CPA_NonEq2022}. However, we summarize it here for the purpose of a reasonably self-contained text. The algorithm starts with an initial guess of the hybridization function $\Delta(t,t')$ describing the effective medium to which the impurity is coupled. This hybridization is used to calculate the noninterating Green's function $\mathcal{G}_{V_i}(t,t')$ for a disorder configuration $V_i$ on the impurity:
\begin{equation}
 \label{eq:Gscript}
 \mathcal{G}_{V_i}(t,t') = \left( \left(i\partial_t + \mu -V_i \right)\delta_c - \Delta) \right)^{-1}(t,t')
\end{equation}
This noninteracting Green's function is used to compute the interacting Green's function $G_{V_i}(t,t')$ via the Dyson equation once the self-energy $\Sigma_{v_i}(t,t')$ has been calculated:
\begin{equation}
 \label{eq:ImpurityG}
 G_{V_i}(t,t')=  \left[ \mathcal{G}_{V_i}^{-1} - \Sigma_{V_i} \right]^{-1}(t,t').
\end{equation}
Here, we use second order perturbation theory as an impurity solver and the self-energy is obtained by:
\begin{equation}
\label{eq:sigmaU} 
 \Sigma_{V_i}(t,t') = -U(t)U(t') \mathcal{G}_{V_i}(t,t')^2 \mathcal{G}_{V_i}(t',t).
\end{equation}
The disorder averaged Green's function $G_{ave}(t,t')$ is obtained by averaging the Green's function $G_{V_i}(t,t')$ on all disorder configurations:
\begin{equation}
    G_{ave}(t,t') = \langle G_{V_i}(t,t') \rangle_{\{V\}}.
\end{equation}
For binary disorder, this average involves two disorder configurations: $V_i = W/2$ and $V_i = -W/2$, both with probability $1/2$. From the disorder averaged Green's function, a new Hybridization function is obtained via:
\begin{equation}
 \label{eq:hybridization}
 \Delta(t,t') = t^{*^2} G_{ave}(t,t').
\end{equation}
and the cycle is repeated with the calculation of the updated impurity non-interacting Green's function, until convergence is obtained within some desired criterion on the disorder averaged Green's function. 

It is clear that this self-consistency loop amounts to the nonequilibrium DMFT for the clean (disorder-free) system and to nonequilibrium CPA for the disordered noninteracting system. It effectively constructs an effective medium that captures both the effect of electron-electron interaction and the effect of disorder for the nonequilibrium system.

The forward and the backward horizontal parts of the contour are each discretized in $N_t = (t_{max} - t_{min})/\Delta t$ time steps;  where $\Delta t$ is the finite time step. The vertical branch is divided into $N_{\tau}$ time steps. This way, the contour-ordered Green's function is represented by a matrix of size $(2N_t + N_{\tau}) \times (2N_t + N_{\tau})$ and we use the matrix formalism described in Ref.[\onlinecite{49a}]. Alternative approaches have been formulated with equations in terms of the different types of Green's functions~\cite{50a}. In the calculations presented in this paper, we use $N_{\tau} = 100$.
Also, the calculations are carried out for three different grid sizes and an extrapolation to the $\Delta t \to 0$ is carried out using Lagarange interpolation to quadratic order with typical values of $N_t \sim 1000$.

\section{Results}
\label{sec:results}

\subsection{Equilibrium density of states}

\noindent The equilibrium density of states is obtained by running the algorithm for the Hamiltonian with fixed disorder strength and interaction strength. The density of states is then calculated by first switching the retarded Green's function from the $(t,t')$ real time coordinates to the $(T_{ave}, t_{rel})$ Wigner coordinates. Where $T_{ave} = (t+t')/2$ and $t_{rel} = t-t'$. For the equilibrium system, this Green's function has no dependence on $T_{ave}$. However, for the purpose of being able to perform a more accurate numerical Fourier transform, we pick $T_{ave}$ that is half of its maximum value, allowing the longest range of relative time, and we perform the Fourier transform with respect to $t_{rel}$ to obtain $G^R(T_{ave}, \omega)$. The density of states is then given by:
\begin{equation}
 \rho(\omega) = -\frac{1}{\pi}\mathrm{Im}\left[G^R(\omega)\right].
\end{equation}
Where we have dropped the $T_{ave}$ variable. The densities of states presented below use $t_{min} = -5 $ and $t_{max} = 20$ and the temperature is set to $\beta = 40$.

\begin{figure}[t] 
\includegraphics[width=8.40cm]{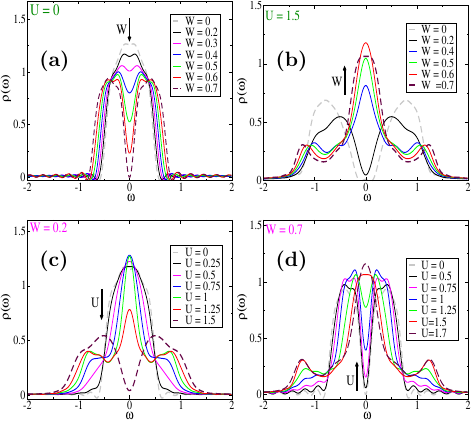}
\caption{Equilibrium density of states of the Anderson-Hubbard model with $U = 0$ (Anderson model) (a), $U = 1.5$ (b) for disorder strength $W$ varying from $0$ to $0.7$. Equilibrium density of states of the Anderson-Hubbard model with $W = 0.2$  (c) and $W = 0.7$ (d) for interaction strength $U$ varying from $0$ to $1.5$ and from $0$ to $1.7$ respectively. The arrows indicate increasing disorder strength in (a) and (b), and increasing interaction strength in (c) and (d).
}
\label{fig:eq_DOS_2X2}
\end{figure}

The possibility of both the disorder and the Mott gap in the Anderson-Hubbard model for a random binary alloy gives rise in the DMFT+CPA calculation to a more complex behavior of the density of states. Figs.~\ref{fig:eq_DOS_2X2} (a) and (b) show the density of states for the noninteracting system ($U = 0$) and for $U = 1.5$ respectively, with disorder strengths varying between $0$ and $0.7$. For the noninteracting system, in the absence of disorder, we recover the semi-elliptical density of states. Note however, the presence of the spurious oscillations due to the truncation of the time axis in the numerical implementation of the formalism on the discretized Keldysh contour. Nevertheless, as seen below, important aspects of the spectrum are captured without resorting to analytical continuation.
Increasing the disorder strength on this noninteracting system gradually splits the spectral weight into two symmetric peaks centered around $\pm W/2$. At finite but small disorder, these are two small bumps and they eventually form a fully gaped spectrum at $W \sim 0.7$ (Fig.~\ref{fig:eq_DOS_2X2} (a)).
A weak but finite interaction strength delays the formation of the disorder gap (not shown). When the interaction is above the critical value of the Mott metal-to-insulator transition for the clean system, as shown in Fig.~\ref{fig:eq_DOS_2X2} (b) for $U = 1.5$, increasing the disorder strength gradually increases the spectral weight around the Fermi energy and eventually closes the gap. 

Figs.~\ref{fig:eq_DOS_2X2} (c) and (d) show the density of states for the $W = 0.2$ (c) and for $W = 0.7$ (d) respectively, with interaction strength varying between $0$ and $1.7$. For $W=0.2$, the density of states is similar to that of the clean system with the opening of the Mott gap delayed to a slightly stronger interaction. This delayed opening of the gap is more apparent for $W= 0.7$ where increasing the interaction strength, is seen to fill the disorder gap. Continuing to increase the interaction eventually leads to the opening of the Mott gap with two Hubbard bands that are each additionally split into two parts due to the presence of the strong binary disorder (not shown). Note however, that this opening of the Mott gap, with moderate-to-strong disorder strength, occurs close to the breakdown of our solution that uses second order perturbation theory as an impurity solver. This breakdown of the nonequilibrium formalism with second order perturbation theory as an impurity solver has been previously documented in the Hubbard model~\cite{56a}.
Similarly, it is well known that important features of disordered systems are missed by the coherent potential approximation in the strong disorder regime. Thus, exploring the system beyond moderate interaction strength and moderate disorder would extend our algorithm outside of its range of validity, and we reserve these explorations for future studies with more accurate solvers.

\subsection{Nonequilibrium dynamics after the interaction quench}
\noindent To probe the interplay of interaction and disorder in the nonequilibrium dynamics, we track the time dependence of the kinetic and potential energy in the system across the interaction quench whereby, with the system initially in equilibrium at a given temperature and with a fixed disorder strength, the interaction is abruptly switched from $U_1 = 0$ to a finite value $U_2$ at which it is subsequently kept constant. We also evaluate the total energy in the system. It is obtained by summing up the kinetic energy and the potential energy~.

For the lattice model, the kinetic energy per lattice site is defined by ~\cite{DMFT_CPA_NonEq2022, 56a}:
\begin{equation}
E_{kin}(t) = \frac{1}{N} \sum_{k, {\sigma}} \epsilon_k \langle c^{\dagger}_{k, \sigma}(t) c_{k, \sigma}(t) \rangle. 
\end{equation}
$N$ is the number of sites, $k$ is the momentum vector, and $\epsilon_k$ is the dispersion relation. Replacing the lattice sum by the integral over band energies, the kinetic energy can thus be rewritten as:
\begin{equation}
E_{kin}(t) = 2\int \rho(\epsilon) \epsilon G^<_{\epsilon}(t,t) d \epsilon.
\end{equation}
Where $\epsilon$ is the band energy. The potential energy follows from the expression of the double occupancy:
\begin{equation}
E_{pot}(t) = [G_{ave}*\Sigma_{ave}]^<(t,t) + \frac{U(t)}{4}.  
\label{eq:Epot}
\end{equation}
$G^<_\epsilon(t,t') $ defined by:
\begin{equation}
G^<_\epsilon(t,t') = \left\{ [(i\partial_t - \epsilon) \delta_c - \Sigma_{ave}]^{-1} \right\}^<(t,t')
\label{eq:GlesserLattice}
\end{equation}
is the lattice lesser Green's function.  $G_{ave}$ is the Green's function averaged over all disorder configurations and $\Sigma_{ave}$ is the self-energy obtained from the Dyson equation with $G_{ave}$ and the noninteracting Green's function. This self-energy thus includes the effects of both the interaction and the disorder. In Equation (\ref{eq:Epot}) for the potential energy,  the lesser part of the convolution  $G_{ave}*\Sigma_{ave}$ is used.

\begin{figure}[t] 
\includegraphics[width=8.40cm]{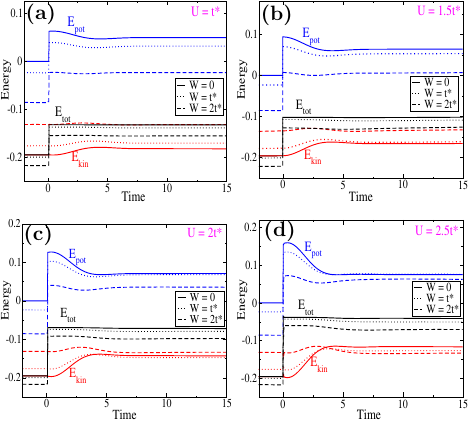}
\caption{Relaxation of the potential (blue), kinetic (red) and total (black) energy as a function of time when the system is quenched at time $t = 0$ from $U_1 = 0$ to $U_2 = t^*$ (a), $U_2 = 1.5 t^*$ (b), $U_2 = 2 t^*$ (c), $U_2 = 2.5 t^*$ (d). Each panel shows the time evolution of the three types of energy for disorder strength $W = 0$ (solid line), $W = t^*$ (dotted line) and $W = 2t^*$ (dashed line).
}
\label{fig:energy_time}
\end{figure}

\begin{figure}[t] 
\begin{center}
\includegraphics[width=7.60cm]{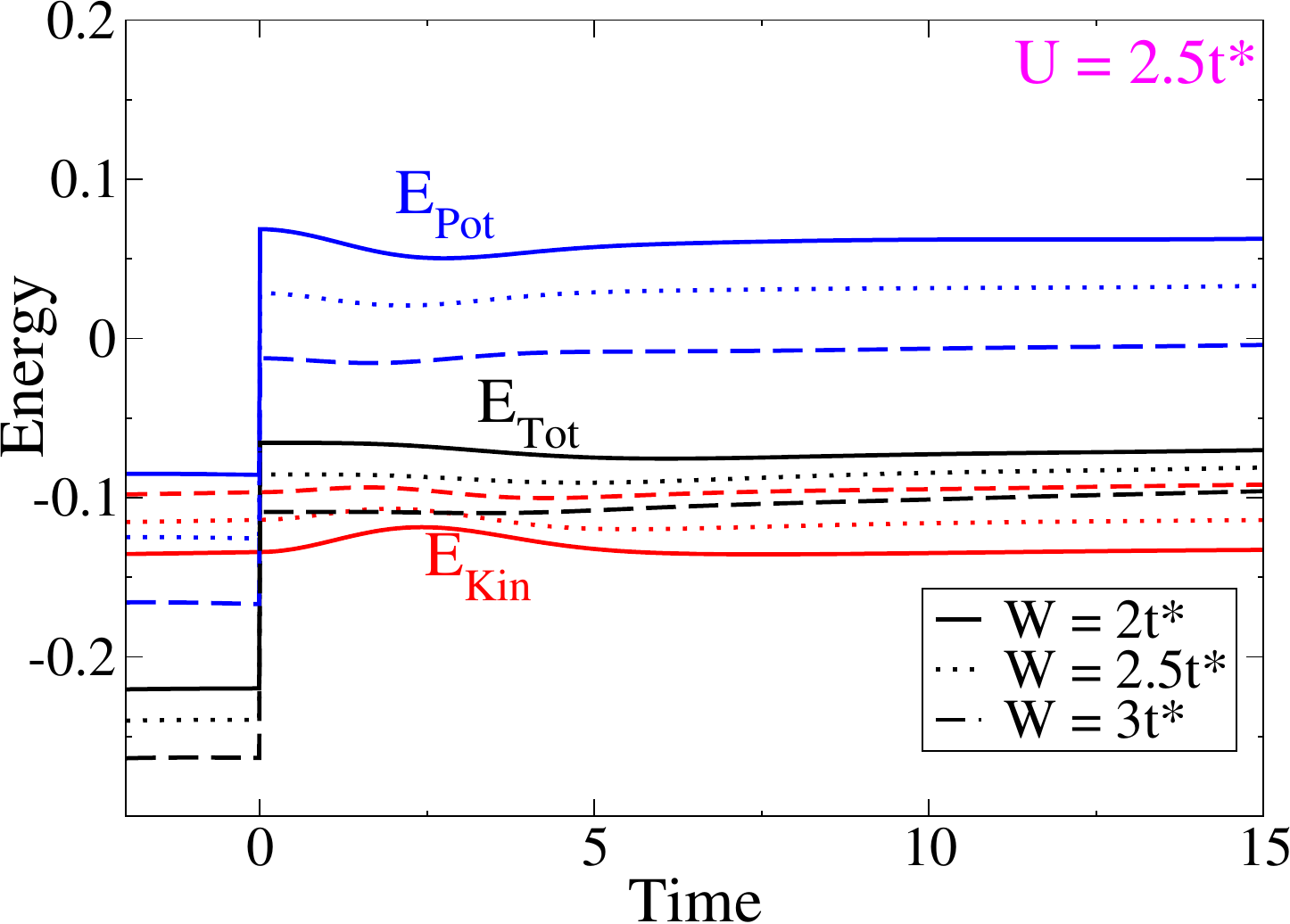}
\vspace{-3mm}
\end{center}
\caption{Relaxation of the potential (blue), kinetic (red) and total (black) energy as a function of time when the system is quenched at time $t = 0$ from $U_1 = 0$ to $U_2 = 2.5 t^*$ as a function of time. The figure presents the trend in the energies for disorder strengths above values studied in Fig.~\ref{fig:energy_time},
$W = 2t^*$ (dotted line), $W = 2.5t^*$ (dashed line) and $W = 3t^*$ (dotted line).
}
\label{fig:energy_time_ModW}
\end{figure}

Fig.~\ref{fig:energy_time} presents the relaxation of the potential (blue), kinetic (red), and total (black) energy as a function of time when the system is quenched at time $t = 0$ from $U_1 = 0$ to $U_2 = t^*$ (a), $U_2 = 1.5 t^*$ (b), $U_2 = 2 t^*$ (c), $U_2 = 2.5 t^*$ (d). Each panel shows the time evolution of the three types of energy for disorder strengths $W = 0$ (solid line), $W = t^*$ (dotted line) and $W = 2t^*$ (dashed line).  After the interaction quench, the total energy changes from an initial value to another value at which it is subsequently kept constant since the system is isolated. This total energy decreases with increased disorder strength. During the transient, the kinetic energy and the potential energy evolve in a complementary manner so as to keep the total energy constant. Immediately across the interaction quench, the change in potential energy for a clean system increases with the final interaction strength. As the disorder strength is increased, this change in potential energy is decreased. After the short-time transient, both the kinetic energy and the potential energy settle into their long-time values. For weak interaction, the long-time value of the kinetic energy increases with the disorder strength. At moderate interaction (Fig.~\ref{fig:energy_time} (d)), this trend is reversed and the long-time value of the kinetic energy decreases with increasing disorder strength. Fig.~\ref{fig:energy_time_ModW} shows the dynamics of the energies after this shifted trend in the kinetic energy with respect to the disorder trend for $U_2 = 2.5 t^*$. Here, the energies are plotted for $W = 2t^*$ (solid line), $W = 2.5t^*$ (dotted line) and $W = 3t^*$ (dashed line). This graph further illustrates the decrease in the change in kinetic energy with increasing disorder strength across the interaction quench.

\begin{figure}[t] 
  \begin{center}
  \setlength{\unitlength}{1cm}
  \begin{picture}(8.0,12.20)(0,0)
    \put(0.0,6.20){\includegraphics*[width=8.0cm, height=6.0cm]{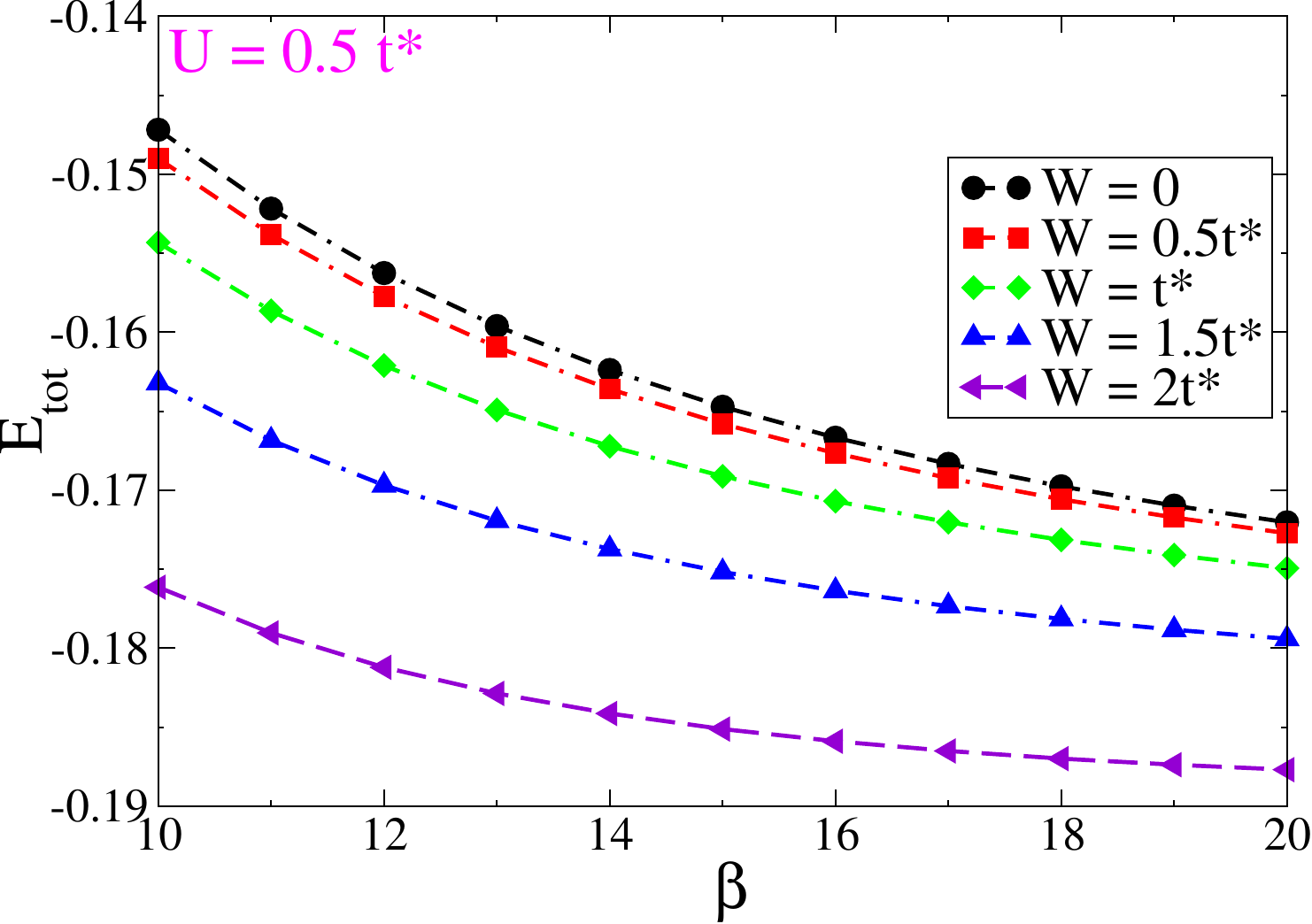}}
    \put(0.0,0.0){\includegraphics*[width=8.0cm, height=6.0cm]{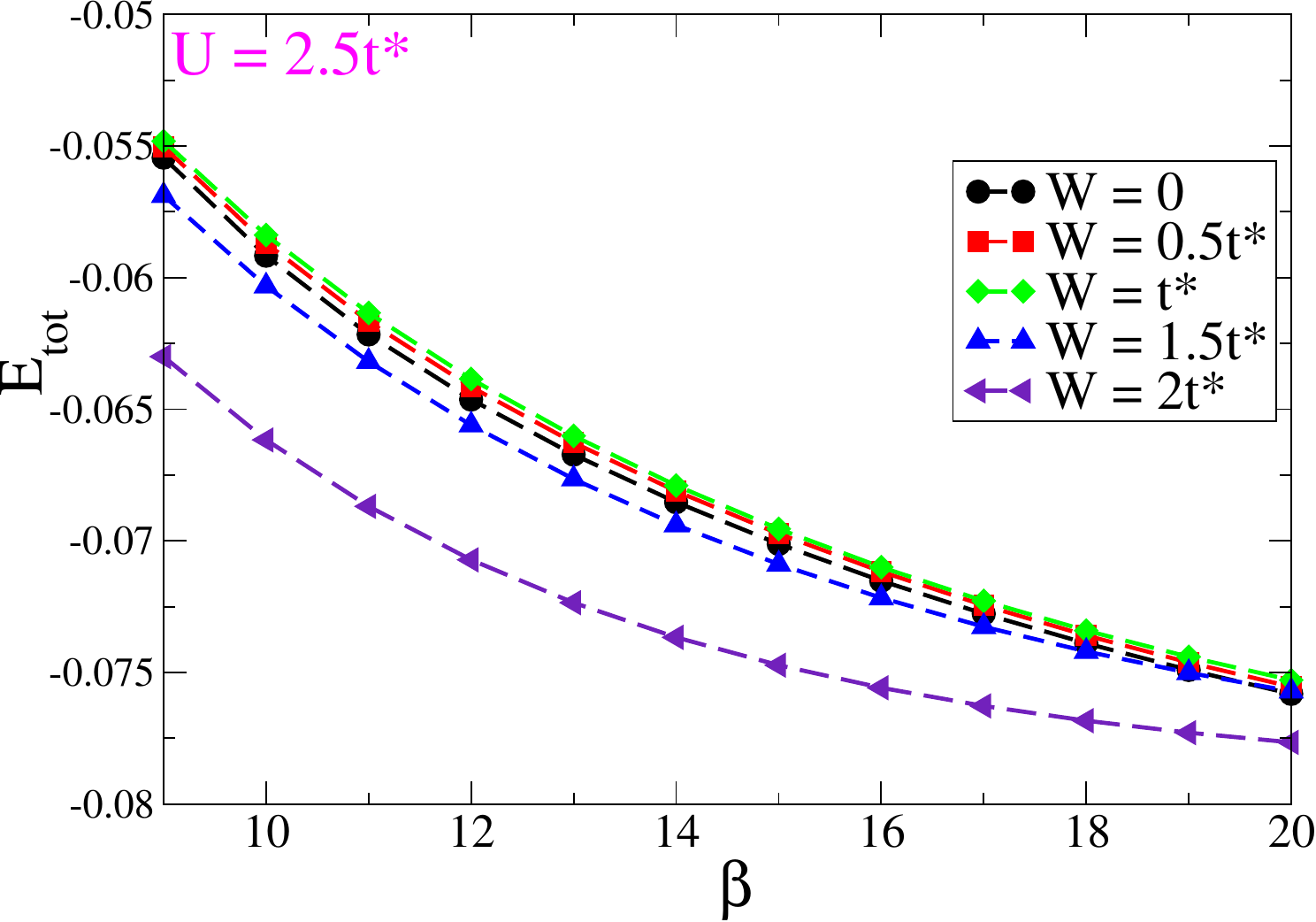}}    
    \end{picture} 
    \vspace{-3mm}
    \end{center}
\caption{Equilibrium total energy as a function of inverse temperature in the interacting binary alloy with $U = 0.5t^*$ (top) and $U = 2.5t^*$(bottom) for disorder strengths $W = 0$ (black circles), $W = 0.5t^*$ (red squares), $W = t^*$ (green  lozenges), $W = 1.5t^*$ (blue triangles), $W = 2t^*$ (purple triangles). The total energy is obtained by summing up the kinetic energy and the potential energy calculated with our nonequilibrium algorithm for the equilibrium system.}  
\label{fig:eq_Etot_beta}
\end{figure}

\begin{figure}[t] 
\includegraphics[width=8.40cm]{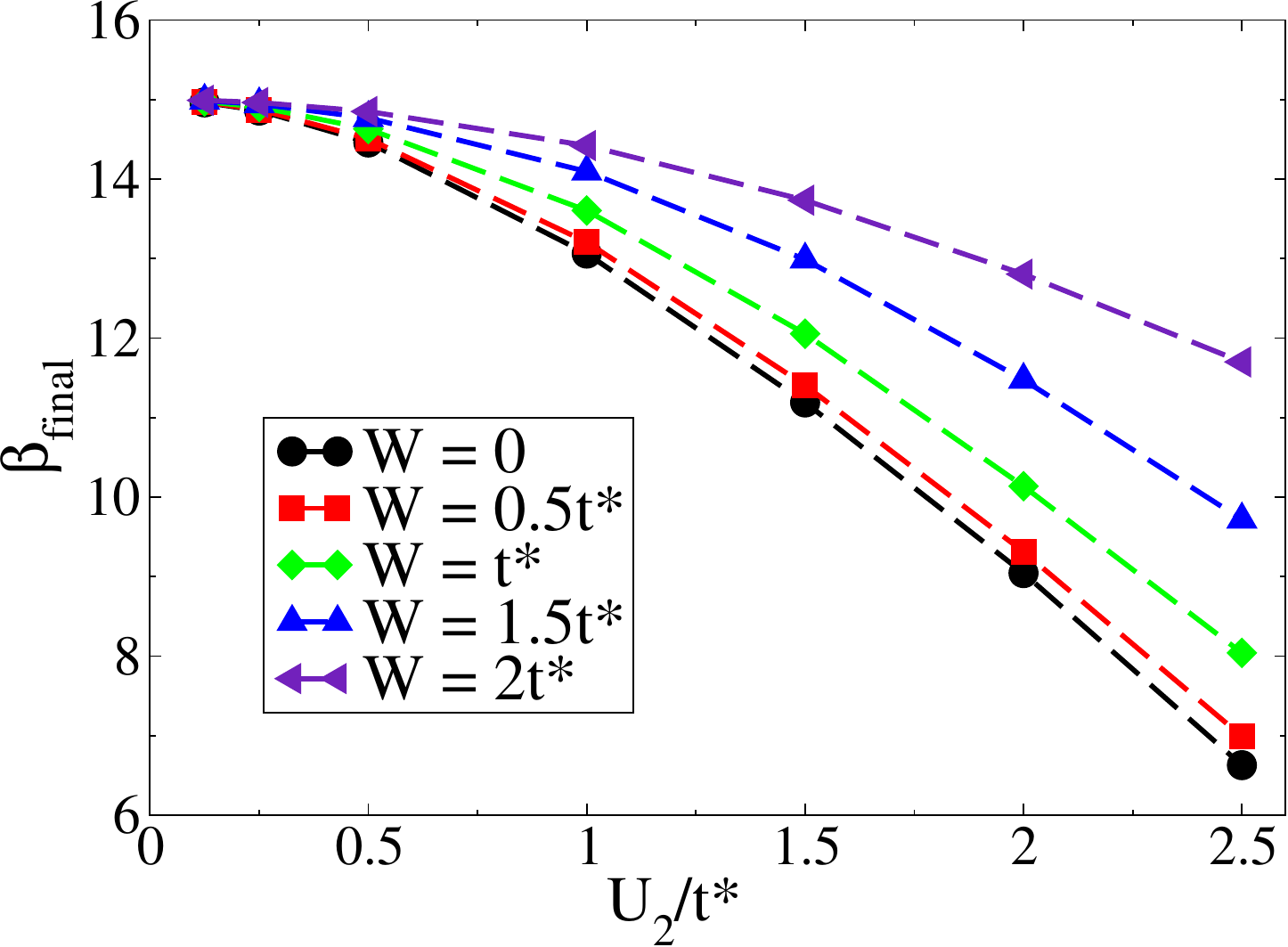}
\caption{Effective temperature as a function of final interaction strength ($U_2$) for different disorder strengths ($W$). The noninteracting system is initially in equilibrium at $\beta = 15$ and the final temperature is obtained by interpolating the total energy of the nonequilibrium system with the equilibrium data as function of inverse temperature for the same interaction and disorder strength.
}
\label{fig:betaEff_U2}
\end{figure}

\subsection{Long-time temperature}
\noindent Before the interaction quench, the system is initially in equilibrium at a given temperature $T= 1/\beta$ with its disorder strength ($W$) and its initial interaction strength (here, $U_1 = 0$). Upon changing, the interaction from $U_1 = 0$ to $U_2$ while keeping the disorder strength constant, the system, as seen through the kinetic energy and the potential energy graphs, first undergoes a nontrivial transient but then settles into a long-time state for the remaining time of the simulation. In earlier studies with ``box" disorder, it was shown by evaluating the distribution function that this long-time state can be accurately considered as thermal~\cite{DMFT_CPA_NonEq2022}. A similar analysis of the distribution function confirms that this long-time state can also be considered thermal. Here, we aim to assess how the temperature for the long-time state  of the binary alloy depends on both the disorder strength and the final interaction strength. The extraction of the temperature via the fluctuation dissipation theorem relating, in a thermal state, the retarded Green's function, the lesser Green's function, and the distribution function is impeded by the presence of a gap in the density of states that renders the numerical procedure unstable. Furthermore, the fluctuation dissipation theorem analysis is carried out in the frequency domain and thus relies on the Fourier transform of both the retarded and the lesser Green's functions. Given the numerical artifacts that arise from the truncation of the time axis at a maximal value in addition to the finite timestep discretization of the Keldysh contour, this procedure is less accurate for low interaction and disorder, and in general, when the system has sharp band edges that lead to cumbersome Gibbs phenomenon. Here, we instead utilize the equilibrium \textit{energy versus temperature} data to infer the effective temperature of the nonequilibrium system when it has settled into its long-time state. Namely, we use the total energy of the system with a given disorder strength after the interaction quench from the noninteracting system onto a given final interaction, and interpolate with the equilibrium total energy data as a function of inverse temperature for the same interaction and the same disorder strength, to extract the longtime temperature of the system.

Fig.~\ref{fig:eq_Etot_beta} shows the equilibrium total energy as a function of inverse temperature for $U = 0.5t^*$ (top) and $U = 2.5t^*$ (bottom), for disorder strengths $W = 0$ (black circles), $W = 0.5t^*$ (red squares), $W = t^*$ (green  lozenges), $W = 1.5t^*$ (blue triangles), and $W = 2t^*$ (purple triangles). This total energy is obtained by summing up the kinetic energy and the potential energy calculated with our time domain algorithm for the equilibrium system. Figure~\ref{fig:betaEff_U2} shows the effective temperature as a function of the final interaction strength ($U_2$) for disorder strengths $W = 0$ (black circles), $W = 0.5t^*$ (red squares), $W = t^*$ (green lozenges), $W = 1.5t^*$ (blue triangles), and $W = 2t^*$ (purple triangles). The noninteracting system is initially in equilibrium at $\beta = 15$ before the interaction quench. 
The graph shows that, for a given disorder strength, the long-time effective temperature of the system, after the transient following the interaction quench, increases monotonically in a nonlinear manner as a function of the final interaction strength. In addition, for a given final interaction, we see that this long-time effective temperature can be tuned over a fairly broad range by the disorder strength, with stronger disorder resulting in a smaller change in the final temperature of the system. This behavior is consistent with the reduced change in the kinetic energy or in the potential energy seen in Fig.~\ref{fig:energy_time} and Fig.~\ref{fig:energy_time_ModW}.

\section{Conclusion}
\label{sec:conclusion}
\noindent We have used the nonequilibrium DMFT+CPA embedding scheme to investigated  the nonequilibrium dynamics of a binary disordered alloy when it is subjected to an interaction quench. The method combines the capacity of DMFT to treat strongly correlated systems and the capacity of CPA to treat disordered systems, to effectively address the interplay of disorder and interaction for the nonequilibrium system. Our system is described by the Anderson-Hubbard model and the disorder is described by a binary distribution of site energies. We used the nonequilibrium method to calculate the equilibrium density of states of the disordered correlated binary alloy. This density of states shows the alteration of the disorder-driven gap by the interaction, and the modification of the interaction-driven Mott gap by the disorder. We assess the relaxation of the system, initially in equilibrium at a given temperature, when the interaction strength is abruptly changed from zero to a finite value at which it is subsequently kept constant. By tracking the evolution in time of the total, kinetic and potential energy across the interaction quench, we find that binary disorder affects the relaxation of the system in a significant manner depending on the final interaction strength after the quench. In addition, we calculate the effective temperature of the system after it has gone through its initial nontrivial transient phase and settled into its long-time state. The effective temperature is found to increase with the final interaction strength. However, for a given final interaction, the final temperature can be tuned over a fairly broad range by the disorder strength. The results provide important insights into the interplay, away from equilibrium, between disorder and interaction for a binary alloy with a method that effectively captures the possibility in the system of two types of insulating phases.

\section*{Acknowledgments}
\noindent We acknowledge useful conversations with Hanna Terletska, Ka-Ming Tam, Jong Han and Enrico Arrigoni. This work was supported by the Department of Energy, Office of Science, Basic Energy Sciences, under grant number DE-SC0024139.\\

\bibliography{mainBib}

\begin{thebibliography}{52}%
\makeatletter
\providecommand \@ifxundefined [1]{%
 \@ifx{#1\undefined}
}%
\providecommand \@ifnum [1]{%
 \ifnum #1\expandafter \@firstoftwo
 \else \expandafter \@secondoftwo
 \fi
}%
\providecommand \@ifx [1]{%
 \ifx #1\expandafter \@firstoftwo
 \else \expandafter \@secondoftwo
 \fi
}%
\providecommand \natexlab [1]{#1}%
\providecommand \enquote  [1]{``#1''}%
\providecommand \bibnamefont  [1]{#1}%
\providecommand \bibfnamefont [1]{#1}%
\providecommand \citenamefont [1]{#1}%
\providecommand \href@noop [0]{\@secondoftwo}%
\providecommand \href [0]{\begingroup \@sanitize@url \@href}%
\providecommand \@href[1]{\@@startlink{#1}\@@href}%
\providecommand \@@href[1]{\endgroup#1\@@endlink}%
\providecommand \@sanitize@url [0]{\catcode `\\12\catcode `\$12\catcode
  `\&12\catcode `\#12\catcode `\^12\catcode `\_12\catcode `\%12\relax}%
\providecommand \@@startlink[1]{}%
\providecommand \@@endlink[0]{}%
\providecommand \url  [0]{\begingroup\@sanitize@url \@url }%
\providecommand \@url [1]{\endgroup\@href {#1}{\urlprefix }}%
\providecommand \urlprefix  [0]{URL }%
\providecommand \Eprint [0]{\href }%
\providecommand \doibase [0]{https://doi.org/}%
\providecommand \selectlanguage [0]{\@gobble}%
\providecommand \bibinfo  [0]{\@secondoftwo}%
\providecommand \bibfield  [0]{\@secondoftwo}%
\providecommand \translation [1]{[#1]}%
\providecommand \BibitemOpen [0]{}%
\providecommand \bibitemStop [0]{}%
\providecommand \bibitemNoStop [0]{.\EOS\space}%
\providecommand \EOS [0]{\spacefactor3000\relax}%
\providecommand \BibitemShut  [1]{\csname bibitem#1\endcsname}%
\let\auto@bib@innerbib\@empty
\bibitem [{\citenamefont {Anderson}(1958)}]{AndersonPhysRev1958}%
  \BibitemOpen
  \bibfield  {author} {\bibinfo {author} {\bibfnamefont {P.~W.}\ \bibnamefont
  {Anderson}},\ }\bibfield  {title} {\bibinfo {title} {Absence of diffusion in
  certain random lattices},\ }\href {https://doi.org/10.1103/PhysRev.109.1492}
  {\bibfield  {journal} {\bibinfo  {journal} {Phys. Rev.}\ }\textbf {\bibinfo
  {volume} {109}},\ \bibinfo {pages} {1492} (\bibinfo {year}
  {1958})}\BibitemShut {NoStop}%
\bibitem [{\citenamefont {Lee}\ and\ \citenamefont
  {Ramakrishnan}(1985)}]{LeeRamakrishnanRevModPhys1985}%
  \BibitemOpen
  \bibfield  {author} {\bibinfo {author} {\bibfnamefont {P.~A.}\ \bibnamefont
  {Lee}}\ and\ \bibinfo {author} {\bibfnamefont {T.~V.}\ \bibnamefont
  {Ramakrishnan}},\ }\bibfield  {title} {\bibinfo {title} {Disordered
  electronic systems},\ }\href {https://doi.org/10.1103/RevModPhys.57.287}
  {\bibfield  {journal} {\bibinfo  {journal} {Rev. Mod. Phys.}\ }\textbf
  {\bibinfo {volume} {57}},\ \bibinfo {pages} {287} (\bibinfo {year}
  {1985})}\BibitemShut {NoStop}%
\bibitem [{\citenamefont {Abrahams}(2010)}]{ElihuBook_50years_of_AL_2010}%
  \BibitemOpen
  \bibfield  {author} {\bibinfo {author} {\bibfnamefont {E.}~\bibnamefont
  {Abrahams}},\ }\href {https://doi.org/10.1142/7663} {\emph {\bibinfo {title}
  {50 Years of Anderson Localization}}}\ (\bibinfo  {publisher} {WORLD
  SCIENTIFIC},\ \bibinfo {year} {2010})\ \Eprint
  {https://arxiv.org/abs/https://www.worldscientific.com/doi/pdf/10.1142/7663}
  {https://www.worldscientific.com/doi/pdf/10.1142/7663} \BibitemShut {NoStop}%
\bibitem [{\citenamefont {and}(1967)}]{MottAdvPhys1967}%
  \BibitemOpen
  \bibfield  {author} {\bibinfo {author} {\bibfnamefont {N.~M.}\ \bibnamefont
  {and}},\ }\bibfield  {title} {\bibinfo {title} {Electrons in disordered
  structures},\ }\href {https://doi.org/10.1080/00018736700101265} {\bibfield
  {journal} {\bibinfo  {journal} {Advances in Physics}\ }\textbf {\bibinfo
  {volume} {16}},\ \bibinfo {pages} {49} (\bibinfo {year} {1967})},\ \Eprint
  {https://arxiv.org/abs/https://doi.org/10.1080/00018736700101265}
  {https://doi.org/10.1080/00018736700101265} \BibitemShut {NoStop}%
\bibitem [{\citenamefont {Thouless}(1974)}]{THOULESS1974}%
  \BibitemOpen
  \bibfield  {author} {\bibinfo {author} {\bibfnamefont {D.}~\bibnamefont
  {Thouless}},\ }\bibfield  {title} {\bibinfo {title} {Electrons in disordered
  systems and the theory of localization},\ }\href
  {https://doi.org/https://doi.org/10.1016/0370-1573(74)90029-5} {\bibfield
  {journal} {\bibinfo  {journal} {Physics Reports}\ }\textbf {\bibinfo {volume}
  {13}},\ \bibinfo {pages} {93} (\bibinfo {year} {1974})}\BibitemShut {NoStop}%
\bibitem [{\citenamefont {Kirkpatrick}(1985)}]{KirkpatrickPRB1985}%
  \BibitemOpen
  \bibfield  {author} {\bibinfo {author} {\bibfnamefont {T.~R.}\ \bibnamefont
  {Kirkpatrick}},\ }\bibfield  {title} {\bibinfo {title} {Localization of
  acoustic waves},\ }\href {https://doi.org/10.1103/PhysRevB.31.5746}
  {\bibfield  {journal} {\bibinfo  {journal} {Phys. Rev. B}\ }\textbf {\bibinfo
  {volume} {31}},\ \bibinfo {pages} {5746} (\bibinfo {year}
  {1985})}\BibitemShut {NoStop}%
\bibitem [{\citenamefont {Lagendijk}\ \emph {et~al.}(2009)\citenamefont
  {Lagendijk}, \citenamefont {Tiggelen},\ and\ \citenamefont
  {Wiersma}}]{LagendijkPhysToday2009}%
  \BibitemOpen
  \bibfield  {author} {\bibinfo {author} {\bibfnamefont {A.}~\bibnamefont
  {Lagendijk}}, \bibinfo {author} {\bibfnamefont {B.~v.}\ \bibnamefont
  {Tiggelen}},\ and\ \bibinfo {author} {\bibfnamefont {D.~S.}\ \bibnamefont
  {Wiersma}},\ }\bibfield  {title} {\bibinfo {title} {Fifty years of anderson
  localization},\ }\href {https://doi.org/10.1063/1.3206091} {\bibfield
  {journal} {\bibinfo  {journal} {Physics Today}\ }\textbf {\bibinfo {volume}
  {62}},\ \bibinfo {pages} {24} (\bibinfo {year} {2009})},\ \Eprint
  {https://arxiv.org/abs/https://pubs.aip.org/physicstoday/article-pdf/62/8/24/11140320/24\_1\_online.pdf}
  {https://pubs.aip.org/physicstoday/article-pdf/62/8/24/11140320/24\_1\_online.pdf}
  \BibitemShut {NoStop}%
\bibitem [{\citenamefont {John}(1984)}]{SajeevPRL1984}%
  \BibitemOpen
  \bibfield  {author} {\bibinfo {author} {\bibfnamefont {S.}~\bibnamefont
  {John}},\ }\bibfield  {title} {\bibinfo {title} {Electromagnetic absorption
  in a disordered medium near a photon mobility edge},\ }\href
  {https://doi.org/10.1103/PhysRevLett.53.2169} {\bibfield  {journal} {\bibinfo
   {journal} {Phys. Rev. Lett.}\ }\textbf {\bibinfo {volume} {53}},\ \bibinfo
  {pages} {2169} (\bibinfo {year} {1984})}\BibitemShut {NoStop}%
\bibitem [{\citenamefont {Wiersma}\ \emph {et~al.}(1997)\citenamefont
  {Wiersma}, \citenamefont {Bartolini}, \citenamefont {Lagendijk},\ and\
  \citenamefont {Righini}}]{Wiersma1997}%
  \BibitemOpen
  \bibfield  {author} {\bibinfo {author} {\bibfnamefont {D.~S.}\ \bibnamefont
  {Wiersma}}, \bibinfo {author} {\bibfnamefont {P.}~\bibnamefont {Bartolini}},
  \bibinfo {author} {\bibfnamefont {A.}~\bibnamefont {Lagendijk}},\ and\
  \bibinfo {author} {\bibfnamefont {R.}~\bibnamefont {Righini}},\ }\bibfield
  {title} {\bibinfo {title} {Localization of light in a disordered medium},\
  }\href {https://doi.org/10.1038/37757} {\bibfield  {journal} {\bibinfo
  {journal} {Nature}\ }\textbf {\bibinfo {volume} {390}},\ \bibinfo {pages}
  {671} (\bibinfo {year} {1997})}\BibitemShut {NoStop}%
\bibitem [{\citenamefont {Segev}\ \emph {et~al.}(2013)\citenamefont {Segev},
  \citenamefont {Silberberg},\ and\ \citenamefont
  {Christodoulides}}]{Segev2013}%
  \BibitemOpen
  \bibfield  {author} {\bibinfo {author} {\bibfnamefont {M.}~\bibnamefont
  {Segev}}, \bibinfo {author} {\bibfnamefont {Y.}~\bibnamefont {Silberberg}},\
  and\ \bibinfo {author} {\bibfnamefont {D.~N.}\ \bibnamefont
  {Christodoulides}},\ }\bibfield  {title} {\bibinfo {title} {Anderson
  localization of light},\ }\href {https://doi.org/10.1038/nphoton.2013.30}
  {\bibfield  {journal} {\bibinfo  {journal} {Nature Photonics}\ }\textbf
  {\bibinfo {volume} {7}},\ \bibinfo {pages} {197} (\bibinfo {year}
  {2013})}\BibitemShut {NoStop}%
\bibitem [{\citenamefont {Kramer}\ and\ \citenamefont {MacKinnon}(1993)}]{1b}%
  \BibitemOpen
  \bibfield  {author} {\bibinfo {author} {\bibfnamefont {B.}~\bibnamefont
  {Kramer}}\ and\ \bibinfo {author} {\bibfnamefont {A.}~\bibnamefont
  {MacKinnon}},\ }\bibfield  {title} {\bibinfo {title} {Localization: theory
  and experiment},\ }\href {https://doi.org/10.1088/0034-4885/56/12/001}
  {\bibfield  {journal} {\bibinfo  {journal} {Reports on Progress in Physics}\
  }\textbf {\bibinfo {volume} {56}},\ \bibinfo {pages} {1469} (\bibinfo {year}
  {1993})}\BibitemShut {NoStop}%
\bibitem [{\citenamefont {Markos}(2006)}]{2b}%
  \BibitemOpen
  \bibfield  {author} {\bibinfo {author} {\bibfnamefont {P.}~\bibnamefont
  {Markos}},\ }\bibfield  {title} {\bibinfo {title} {Numerical analysis of
  anderson localization},\ }\href {doi:10.2478/v10155-010-0081-0} {\bibfield
  {journal} {\bibinfo  {journal} {Acta Phys. Slovaca}\ }\textbf {\bibinfo
  {volume} {56}},\ \bibinfo {pages} {561} (\bibinfo {year} {2006})}\BibitemShut
  {NoStop}%
\bibitem [{\citenamefont {Brandes}\ and\ \citenamefont {Kettemann}(2003)}]{3c}%
  \BibitemOpen
  \bibfield  {author} {\bibinfo {author} {\bibfnamefont {T.}~\bibnamefont
  {Brandes}}\ and\ \bibinfo {author} {\bibfnamefont {S.}~\bibnamefont
  {Kettemann}},\ }\href {https://books.google.com/books?id=EaFjyalWGFMC} {\emph
  {\bibinfo {title} {Anderson Localization and Its Ramifications: Disorder,
  Phase Coherence, and Electron Correlations}}},\ Lecture Notes in Physics\
  (\bibinfo  {publisher} {Springer Berlin Heidelberg},\ \bibinfo {year}
  {2003})\BibitemShut {NoStop}%
\bibitem [{\citenamefont {Elliott}\ \emph {et~al.}(1974)\citenamefont
  {Elliott}, \citenamefont {Krumhansl},\ and\ \citenamefont {Leath}}]{4b}%
  \BibitemOpen
  \bibfield  {author} {\bibinfo {author} {\bibfnamefont {R.~J.}\ \bibnamefont
  {Elliott}}, \bibinfo {author} {\bibfnamefont {J.~A.}\ \bibnamefont
  {Krumhansl}},\ and\ \bibinfo {author} {\bibfnamefont {P.~L.}\ \bibnamefont
  {Leath}},\ }\bibfield  {title} {\bibinfo {title} {The theory and properties
  of randomly disordered crystals and related physical systems},\ }\href
  {https://doi.org/10.1103/RevModPhys.46.465} {\bibfield  {journal} {\bibinfo
  {journal} {Rev. Mod. Phys.}\ }\textbf {\bibinfo {volume} {46}},\ \bibinfo
  {pages} {465} (\bibinfo {year} {1974})}\BibitemShut {NoStop}%
\bibitem [{\citenamefont {Pal}\ and\ \citenamefont
  {Huse}(2010)}]{PalHuse_prb2010}%
  \BibitemOpen
  \bibfield  {author} {\bibinfo {author} {\bibfnamefont {A.}~\bibnamefont
  {Pal}}\ and\ \bibinfo {author} {\bibfnamefont {D.~A.}\ \bibnamefont {Huse}},\
  }\bibfield  {title} {\bibinfo {title} {Many-body localization phase
  transition},\ }\href {https://doi.org/10.1103/PhysRevB.82.174411} {\bibfield
  {journal} {\bibinfo  {journal} {Phys. Rev. B}\ }\textbf {\bibinfo {volume}
  {82}},\ \bibinfo {pages} {174411} (\bibinfo {year} {2010})}\BibitemShut
  {NoStop}%
\bibitem [{\citenamefont {Basko}\ \emph {et~al.}(2006)\citenamefont {Basko},
  \citenamefont {Aleiner},\ and\ \citenamefont {Altshuler}}]{BASKO20061126}%
  \BibitemOpen
  \bibfield  {author} {\bibinfo {author} {\bibfnamefont {D.}~\bibnamefont
  {Basko}}, \bibinfo {author} {\bibfnamefont {I.}~\bibnamefont {Aleiner}},\
  and\ \bibinfo {author} {\bibfnamefont {B.}~\bibnamefont {Altshuler}},\
  }\bibfield  {title} {\bibinfo {title} {Metal–insulator transition in a
  weakly interacting many-electron system with localized single-particle
  states},\ }\href {https://doi.org/https://doi.org/10.1016/j.aop.2005.11.014}
  {\bibfield  {journal} {\bibinfo  {journal} {Annals of Physics}\ }\textbf
  {\bibinfo {volume} {321}},\ \bibinfo {pages} {1126} (\bibinfo {year}
  {2006})}\BibitemShut {NoStop}%
\bibitem [{\citenamefont {Abanin}\ \emph {et~al.}(2019)\citenamefont {Abanin},
  \citenamefont {Altman}, \citenamefont {Bloch},\ and\ \citenamefont
  {Serbyn}}]{AbaninAltmanBloch_RMP2019}%
  \BibitemOpen
  \bibfield  {author} {\bibinfo {author} {\bibfnamefont {D.~A.}\ \bibnamefont
  {Abanin}}, \bibinfo {author} {\bibfnamefont {E.}~\bibnamefont {Altman}},
  \bibinfo {author} {\bibfnamefont {I.}~\bibnamefont {Bloch}},\ and\ \bibinfo
  {author} {\bibfnamefont {M.}~\bibnamefont {Serbyn}},\ }\bibfield  {title}
  {\bibinfo {title} {Colloquium: Many-body localization, thermalization, and
  entanglement},\ }\href {https://doi.org/10.1103/RevModPhys.91.021001}
  {\bibfield  {journal} {\bibinfo  {journal} {Rev. Mod. Phys.}\ }\textbf
  {\bibinfo {volume} {91}},\ \bibinfo {pages} {021001} (\bibinfo {year}
  {2019})}\BibitemShut {NoStop}%
\bibitem [{\citenamefont {Oganesyan}\ and\ \citenamefont
  {Huse}(2007)}]{OganesyanHuse_PRB2007}%
  \BibitemOpen
  \bibfield  {author} {\bibinfo {author} {\bibfnamefont {V.}~\bibnamefont
  {Oganesyan}}\ and\ \bibinfo {author} {\bibfnamefont {D.~A.}\ \bibnamefont
  {Huse}},\ }\bibfield  {title} {\bibinfo {title} {Localization of interacting
  fermions at high temperature},\ }\href
  {https://doi.org/10.1103/PhysRevB.75.155111} {\bibfield  {journal} {\bibinfo
  {journal} {Phys. Rev. B}\ }\textbf {\bibinfo {volume} {75}},\ \bibinfo
  {pages} {155111} (\bibinfo {year} {2007})}\BibitemShut {NoStop}%
\bibitem [{\citenamefont {Nandkishore}\ and\ \citenamefont
  {Huse}(2015)}]{NandkishoreHuseAnRevCondMattPhys2015}%
  \BibitemOpen
  \bibfield  {author} {\bibinfo {author} {\bibfnamefont {R.}~\bibnamefont
  {Nandkishore}}\ and\ \bibinfo {author} {\bibfnamefont {D.~A.}\ \bibnamefont
  {Huse}},\ }\bibfield  {title} {\bibinfo {title} {Many-body localization and
  thermalization in quantum statistical mechanics},\ }\href
  {https://doi.org/https://doi.org/10.1146/annurev-conmatphys-031214-014726}
  {\bibfield  {journal} {\bibinfo  {journal} {Annual Review of Condensed Matter
  Physics}\ }\textbf {\bibinfo {volume} {6}},\ \bibinfo {pages} {15} (\bibinfo
  {year} {2015})}\BibitemShut {NoStop}%
\bibitem [{\citenamefont {Metzner}\ and\ \citenamefont
  {Vollhardt}(1989)}]{33a}%
  \BibitemOpen
  \bibfield  {author} {\bibinfo {author} {\bibfnamefont {W.}~\bibnamefont
  {Metzner}}\ and\ \bibinfo {author} {\bibfnamefont {D.}~\bibnamefont
  {Vollhardt}},\ }\bibfield  {title} {\bibinfo {title} {Correlated lattice
  fermions in $d=\ensuremath{\infty}$ dimensions},\ }\href
  {https://doi.org/10.1103/PhysRevLett.62.324} {\bibfield  {journal} {\bibinfo
  {journal} {Phys. Rev. Lett.}\ }\textbf {\bibinfo {volume} {62}},\ \bibinfo
  {pages} {324} (\bibinfo {year} {1989})}\BibitemShut {NoStop}%
\bibitem [{\citenamefont {Kuramoto}(1985)}]{34a}%
  \BibitemOpen
  \bibfield  {author} {\bibinfo {author} {\bibfnamefont {Y.}~\bibnamefont
  {Kuramoto}},\ }\href@noop {} {\emph {\bibinfo {title} {Springer Series in
  Solid State Science, edited by T. Kasuya and T. Sao}}},\ Vol.~\bibinfo
  {volume} {62}\ (\bibinfo  {publisher} {Springer},\ \bibinfo {year} {1985})\
  p.\ \bibinfo {pages} {152}\BibitemShut {NoStop}%
\bibitem [{\citenamefont {M{\"u}ller-Hartmann}(1989)}]{35a}%
  \BibitemOpen
  \bibfield  {author} {\bibinfo {author} {\bibfnamefont {E.}~\bibnamefont
  {M{\"u}ller-Hartmann}},\ }\bibfield  {title} {\bibinfo {title} {Correlated
  fermions on a lattice in high dimensions},\ }\href
  {https://doi.org/10.1007/BF01311397} {\bibfield  {journal} {\bibinfo
  {journal} {Zeitschrift f{\"u}r Physik B Condensed Matter}\ }\textbf {\bibinfo
  {volume} {74}},\ \bibinfo {pages} {507} (\bibinfo {year} {1989})}\BibitemShut
  {NoStop}%
\bibitem [{\citenamefont {Georges}\ \emph {et~al.}(1996)\citenamefont
  {Georges}, \citenamefont {Kotliar}, \citenamefont {Krauth},\ and\
  \citenamefont {Rozenberg}}]{36a}%
  \BibitemOpen
  \bibfield  {author} {\bibinfo {author} {\bibfnamefont {A.}~\bibnamefont
  {Georges}}, \bibinfo {author} {\bibfnamefont {G.}~\bibnamefont {Kotliar}},
  \bibinfo {author} {\bibfnamefont {W.}~\bibnamefont {Krauth}},\ and\ \bibinfo
  {author} {\bibfnamefont {M.~J.}\ \bibnamefont {Rozenberg}},\ }\bibfield
  {title} {\bibinfo {title} {Dynamical mean-field theory of strongly correlated
  fermion systems and the limit of infinite dimensions},\ }\href
  {https://doi.org/10.1103/RevModPhys.68.13} {\bibfield  {journal} {\bibinfo
  {journal} {Rev. Mod. Phys.}\ }\textbf {\bibinfo {volume} {68}},\ \bibinfo
  {pages} {13} (\bibinfo {year} {1996})}\BibitemShut {NoStop}%
\bibitem [{\citenamefont {Freericks}\ and\ \citenamefont
  {Zlati\ifmmode~\acute{c}\else \'{c}\fi{}}(2003)}]{37a}%
  \BibitemOpen
  \bibfield  {author} {\bibinfo {author} {\bibfnamefont {J.~K.}\ \bibnamefont
  {Freericks}}\ and\ \bibinfo {author} {\bibfnamefont {V.}~\bibnamefont
  {Zlati\ifmmode~\acute{c}\else \'{c}\fi{}}},\ }\bibfield  {title} {\bibinfo
  {title} {Exact dynamical mean-field theory of the falicov-kimball model},\
  }\href {https://doi.org/10.1103/RevModPhys.75.1333} {\bibfield  {journal}
  {\bibinfo  {journal} {Rev. Mod. Phys.}\ }\textbf {\bibinfo {volume} {75}},\
  \bibinfo {pages} {1333} (\bibinfo {year} {2003})}\BibitemShut {NoStop}%
\bibitem [{\citenamefont {Soven}(1967)}]{38a}%
  \BibitemOpen
  \bibfield  {author} {\bibinfo {author} {\bibfnamefont {P.}~\bibnamefont
  {Soven}},\ }\bibfield  {title} {\bibinfo {title} {Coherent-potential model of
  substitutional disordered alloys},\ }\href
  {https://doi.org/10.1103/PhysRev.156.809} {\bibfield  {journal} {\bibinfo
  {journal} {Phys. Rev.}\ }\textbf {\bibinfo {volume} {156}},\ \bibinfo {pages}
  {809} (\bibinfo {year} {1967})}\BibitemShut {NoStop}%
\bibitem [{\citenamefont {Velick\'y}\ \emph {et~al.}(1968)\citenamefont
  {Velick\'y}, \citenamefont {Kirkpatrick},\ and\ \citenamefont
  {Ehrenreich}}]{39a}%
  \BibitemOpen
  \bibfield  {author} {\bibinfo {author} {\bibfnamefont {B.}~\bibnamefont
  {Velick\'y}}, \bibinfo {author} {\bibfnamefont {S.}~\bibnamefont
  {Kirkpatrick}},\ and\ \bibinfo {author} {\bibfnamefont {H.}~\bibnamefont
  {Ehrenreich}},\ }\bibfield  {title} {\bibinfo {title} {Single-site
  approximations in the electronic theory of simple binary alloys},\ }\href
  {https://doi.org/10.1103/PhysRev.175.747} {\bibfield  {journal} {\bibinfo
  {journal} {Phys. Rev.}\ }\textbf {\bibinfo {volume} {175}},\ \bibinfo {pages}
  {747} (\bibinfo {year} {1968})}\BibitemShut {NoStop}%
\bibitem [{\citenamefont {Velick\'y}(1969)}]{40a}%
  \BibitemOpen
  \bibfield  {author} {\bibinfo {author} {\bibfnamefont {B.}~\bibnamefont
  {Velick\'y}},\ }\bibfield  {title} {\bibinfo {title} {Theory of electronic
  transport in disordered binary alloys: Coherent-potential approximation},\
  }\href {https://doi.org/10.1103/PhysRev.184.614} {\bibfield  {journal}
  {\bibinfo  {journal} {Phys. Rev.}\ }\textbf {\bibinfo {volume} {184}},\
  \bibinfo {pages} {614} (\bibinfo {year} {1969})}\BibitemShut {NoStop}%
\bibitem [{\citenamefont {Yonezawa}\ and\ \citenamefont
  {Morigaki}(1973)}]{41a}%
  \BibitemOpen
  \bibfield  {author} {\bibinfo {author} {\bibfnamefont {F.}~\bibnamefont
  {Yonezawa}}\ and\ \bibinfo {author} {\bibfnamefont {K.}~\bibnamefont
  {Morigaki}},\ }\bibfield  {title} {\bibinfo {title} {Coherent potential
  approximation. basic concepts and applications},\ }\href
  {https://academic.oup.com/ptps/article/doi/10.1143/PTPS.53.1/1924416}
  {\bibfield  {journal} {\bibinfo  {journal} {Progress of Theoretical Physics
  Supplement}\ }\textbf {\bibinfo {volume} {53}},\ \bibinfo {pages} {1}
  (\bibinfo {year} {1973})}\BibitemShut {NoStop}%
\bibitem [{\citenamefont {Ulmke}\ \emph {et~al.}(1995)\citenamefont {Ulmke},
  \citenamefont {Jani\ifmmode~\check{s}\else \v{s}\fi{}},\ and\ \citenamefont
  {Vollhardt}}]{UlmkeJanisVollhardt}%
  \BibitemOpen
  \bibfield  {author} {\bibinfo {author} {\bibfnamefont {M.}~\bibnamefont
  {Ulmke}}, \bibinfo {author} {\bibfnamefont {V.}~\bibnamefont
  {Jani\ifmmode~\check{s}\else \v{s}\fi{}}},\ and\ \bibinfo {author}
  {\bibfnamefont {D.}~\bibnamefont {Vollhardt}},\ }\bibfield  {title} {\bibinfo
  {title} {Anderson-hubbard model in infinite dimensions},\ }\href
  {https://doi.org/10.1103/PhysRevB.51.10411} {\bibfield  {journal} {\bibinfo
  {journal} {Phys. Rev. B}\ }\textbf {\bibinfo {volume} {51}},\ \bibinfo
  {pages} {10411} (\bibinfo {year} {1995})}\BibitemShut {NoStop}%
\bibitem [{\citenamefont {Janis\ifmmode~\check{}\else \v{}\fi{}}\ and\
  \citenamefont {Vollhardt}(1992)}]{JanisVollhardt_PRB1992}%
  \BibitemOpen
  \bibfield  {author} {\bibinfo {author} {\bibfnamefont {V.}~\bibnamefont
  {Janis\ifmmode~\check{}\else \v{}\fi{}}}\ and\ \bibinfo {author}
  {\bibfnamefont {D.}~\bibnamefont {Vollhardt}},\ }\bibfield  {title} {\bibinfo
  {title} {Coupling of quantum degrees of freedom in strongly interacting
  disordered electron systems},\ }\href
  {https://doi.org/10.1103/PhysRevB.46.15712} {\bibfield  {journal} {\bibinfo
  {journal} {Phys. Rev. B}\ }\textbf {\bibinfo {volume} {46}},\ \bibinfo
  {pages} {15712} (\bibinfo {year} {1992})}\BibitemShut {NoStop}%
\bibitem [{\citenamefont {Lombardo}\ \emph {et~al.}(2006)\citenamefont
  {Lombardo}, \citenamefont {Hayn},\ and\ \citenamefont
  {Japaridze}}]{LombardoEtAl2006}%
  \BibitemOpen
  \bibfield  {author} {\bibinfo {author} {\bibfnamefont {P.}~\bibnamefont
  {Lombardo}}, \bibinfo {author} {\bibfnamefont {R.}~\bibnamefont {Hayn}},\
  and\ \bibinfo {author} {\bibfnamefont {G.~I.}\ \bibnamefont {Japaridze}},\
  }\bibfield  {title} {\bibinfo {title} {Insulator-metal-insulator transition
  and selective spectral-weight transfer in a disordered strongly correlated
  system},\ }\href {https://doi.org/10.1103/PhysRevB.74.085116} {\bibfield
  {journal} {\bibinfo  {journal} {Phys. Rev. B}\ }\textbf {\bibinfo {volume}
  {74}},\ \bibinfo {pages} {085116} (\bibinfo {year} {2006})}\BibitemShut
  {NoStop}%
\bibitem [{\citenamefont {Laad}\ \emph {et~al.}(2001)\citenamefont {Laad},
  \citenamefont {Craco},\ and\ \citenamefont
  {M\"uller-Hartmann}}]{LaadMHartmannPRN2001}%
  \BibitemOpen
  \bibfield  {author} {\bibinfo {author} {\bibfnamefont {M.~S.}\ \bibnamefont
  {Laad}}, \bibinfo {author} {\bibfnamefont {L.}~\bibnamefont {Craco}},\ and\
  \bibinfo {author} {\bibfnamefont {E.}~\bibnamefont {M\"uller-Hartmann}},\
  }\bibfield  {title} {\bibinfo {title} {Effect of strong correlations and
  static diagonal disorder in the $d=\ensuremath{\infty}$ hubbard model},\
  }\href {https://doi.org/10.1103/PhysRevB.64.195114} {\bibfield  {journal}
  {\bibinfo  {journal} {Phys. Rev. B}\ }\textbf {\bibinfo {volume} {64}},\
  \bibinfo {pages} {195114} (\bibinfo {year} {2001})}\BibitemShut {NoStop}%
\bibitem [{\citenamefont {Weh}\ \emph {et~al.}(2021)\citenamefont {Weh},
  \citenamefont {Zhang}, \citenamefont {\"Ostlin}, \citenamefont {Terletska},
  \citenamefont {Bauernfeind}, \citenamefont {Tam}, \citenamefont {Evertz},
  \citenamefont {Byczuk}, \citenamefont {Vollhardt},\ and\ \citenamefont
  {Chioncel}}]{PRB_DMFT_AndersonHubbard_WehVollhardtEtAl2021}%
  \BibitemOpen
  \bibfield  {author} {\bibinfo {author} {\bibfnamefont {A.}~\bibnamefont
  {Weh}}, \bibinfo {author} {\bibfnamefont {Y.}~\bibnamefont {Zhang}}, \bibinfo
  {author} {\bibfnamefont {A.}~\bibnamefont {\"Ostlin}}, \bibinfo {author}
  {\bibfnamefont {H.}~\bibnamefont {Terletska}}, \bibinfo {author}
  {\bibfnamefont {D.}~\bibnamefont {Bauernfeind}}, \bibinfo {author}
  {\bibfnamefont {K.-M.}\ \bibnamefont {Tam}}, \bibinfo {author} {\bibfnamefont
  {H.~G.}\ \bibnamefont {Evertz}}, \bibinfo {author} {\bibfnamefont
  {K.}~\bibnamefont {Byczuk}}, \bibinfo {author} {\bibfnamefont
  {D.}~\bibnamefont {Vollhardt}},\ and\ \bibinfo {author} {\bibfnamefont
  {L.}~\bibnamefont {Chioncel}},\ }\bibfield  {title} {\bibinfo {title}
  {Dynamical mean-field theory of the anderson-hubbard model with local and
  nonlocal disorder in tensor formulation},\ }\href
  {https://doi.org/10.1103/PhysRevB.104.045127} {\bibfield  {journal} {\bibinfo
   {journal} {Phys. Rev. B}\ }\textbf {\bibinfo {volume} {104}},\ \bibinfo
  {pages} {045127} (\bibinfo {year} {2021})}\BibitemShut {NoStop}%
\bibitem [{\citenamefont {Freericks}\ \emph {et~al.}(2006)\citenamefont
  {Freericks}, \citenamefont {Turkowski},\ and\ \citenamefont
  {Zlati\ifmmode~\acute{c}\else \'{c}\fi{}}}]{48a}%
  \BibitemOpen
  \bibfield  {author} {\bibinfo {author} {\bibfnamefont {J.~K.}\ \bibnamefont
  {Freericks}}, \bibinfo {author} {\bibfnamefont {V.~M.}\ \bibnamefont
  {Turkowski}},\ and\ \bibinfo {author} {\bibfnamefont {V.}~\bibnamefont
  {Zlati\ifmmode~\acute{c}\else \'{c}\fi{}}},\ }\bibfield  {title} {\bibinfo
  {title} {Nonequilibrium dynamical mean-field theory},\ }\href
  {https://doi.org/10.1103/PhysRevLett.97.266408} {\bibfield  {journal}
  {\bibinfo  {journal} {Phys. Rev. Lett.}\ }\textbf {\bibinfo {volume} {97}},\
  \bibinfo {pages} {266408} (\bibinfo {year} {2006})}\BibitemShut {NoStop}%
\bibitem [{\citenamefont {Freericks}(2008)}]{49a}%
  \BibitemOpen
  \bibfield  {author} {\bibinfo {author} {\bibfnamefont {J.~K.}\ \bibnamefont
  {Freericks}},\ }\bibfield  {title} {\bibinfo {title} {Quenching bloch
  oscillations in a strongly correlated material: Nonequilibrium dynamical
  mean-field theory},\ }\href {https://doi.org/10.1103/PhysRevB.77.075109}
  {\bibfield  {journal} {\bibinfo  {journal} {Phys. Rev. B}\ }\textbf {\bibinfo
  {volume} {77}},\ \bibinfo {pages} {075109} (\bibinfo {year}
  {2008})}\BibitemShut {NoStop}%
\bibitem [{\citenamefont {Aoki}\ \emph
  {et~al.}(2014{\natexlab{a}})\citenamefont {Aoki}, \citenamefont {Tsuji},
  \citenamefont {Eckstein}, \citenamefont {Kollar}, \citenamefont {Oka},\ and\
  \citenamefont {Werner}}]{50a}%
  \BibitemOpen
  \bibfield  {author} {\bibinfo {author} {\bibfnamefont {H.}~\bibnamefont
  {Aoki}}, \bibinfo {author} {\bibfnamefont {N.}~\bibnamefont {Tsuji}},
  \bibinfo {author} {\bibfnamefont {M.}~\bibnamefont {Eckstein}}, \bibinfo
  {author} {\bibfnamefont {M.}~\bibnamefont {Kollar}}, \bibinfo {author}
  {\bibfnamefont {T.}~\bibnamefont {Oka}},\ and\ \bibinfo {author}
  {\bibfnamefont {P.}~\bibnamefont {Werner}},\ }\bibfield  {title} {\bibinfo
  {title} {Nonequilibrium dynamical mean-field theory and its applications},\
  }\href {https://doi.org/10.1103/RevModPhys.86.779} {\bibfield  {journal}
  {\bibinfo  {journal} {Rev. Mod. Phys.}\ }\textbf {\bibinfo {volume} {86}},\
  \bibinfo {pages} {779} (\bibinfo {year} {2014}{\natexlab{a}})}\BibitemShut
  {NoStop}%
\bibitem [{\citenamefont {Fotso}\ and\ \citenamefont {Freericks}(2020)}]{54a}%
  \BibitemOpen
  \bibfield  {author} {\bibinfo {author} {\bibfnamefont {H.~F.}\ \bibnamefont
  {Fotso}}\ and\ \bibinfo {author} {\bibfnamefont {J.~K.}\ \bibnamefont
  {Freericks}},\ }\bibfield  {title} {\bibinfo {title} {Characterizing the
  non-equilibrium dynamics of field-driven correlated quantum systems},\ }\href
  {https://doi.org/10.3389/fphy.2020.00324} {\bibfield  {journal} {\bibinfo
  {journal} {Frontiers in Physics}\ }\textbf {\bibinfo {volume} {8}},\ \bibinfo
  {pages} {324} (\bibinfo {year} {2020})}\BibitemShut {NoStop}%
\bibitem [{\citenamefont {Fotso}\ \emph {et~al.}(2014)\citenamefont {Fotso},
  \citenamefont {Mikelsons},\ and\ \citenamefont {Freericks}}]{55a}%
  \BibitemOpen
  \bibfield  {author} {\bibinfo {author} {\bibfnamefont {H.}~\bibnamefont
  {Fotso}}, \bibinfo {author} {\bibfnamefont {K.}~\bibnamefont {Mikelsons}},\
  and\ \bibinfo {author} {\bibfnamefont {J.~K.}\ \bibnamefont {Freericks}},\
  }\bibfield  {title} {\bibinfo {title} {Thermalization of field driven quantum
  systems},\ }\href {https://doi.org/10.1038/srep04699} {\bibfield  {journal}
  {\bibinfo  {journal} {Scientific Reports}\ }\textbf {\bibinfo {volume} {4}},\
  \bibinfo {pages} {4699} (\bibinfo {year} {2014})}\BibitemShut {NoStop}%
\bibitem [{\citenamefont {Eckstein}\ \emph {et~al.}(2010)\citenamefont
  {Eckstein}, \citenamefont {Kollar},\ and\ \citenamefont {Werner}}]{56a}%
  \BibitemOpen
  \bibfield  {author} {\bibinfo {author} {\bibfnamefont {M.}~\bibnamefont
  {Eckstein}}, \bibinfo {author} {\bibfnamefont {M.}~\bibnamefont {Kollar}},\
  and\ \bibinfo {author} {\bibfnamefont {P.}~\bibnamefont {Werner}},\
  }\bibfield  {title} {\bibinfo {title} {Interaction quench in the hubbard
  model: Relaxation of the spectral function and the optical conductivity},\
  }\href {https://doi.org/10.1103/PhysRevB.81.115131} {\bibfield  {journal}
  {\bibinfo  {journal} {Phys. Rev. B}\ }\textbf {\bibinfo {volume} {81}},\
  \bibinfo {pages} {115131} (\bibinfo {year} {2010})}\BibitemShut {NoStop}%
\bibitem [{\citenamefont {Eckstein}\ \emph {et~al.}(2009)\citenamefont
  {Eckstein}, \citenamefont {Kollar},\ and\ \citenamefont {Werner}}]{68a}%
  \BibitemOpen
  \bibfield  {author} {\bibinfo {author} {\bibfnamefont {M.}~\bibnamefont
  {Eckstein}}, \bibinfo {author} {\bibfnamefont {M.}~\bibnamefont {Kollar}},\
  and\ \bibinfo {author} {\bibfnamefont {P.}~\bibnamefont {Werner}},\
  }\bibfield  {title} {\bibinfo {title} {Thermalization after an interaction
  quench in the hubbard model},\ }\href
  {https://doi.org/10.1103/PhysRevLett.103.056403} {\bibfield  {journal}
  {\bibinfo  {journal} {Phys. Rev. Lett.}\ }\textbf {\bibinfo {volume} {103}},\
  \bibinfo {pages} {056403} (\bibinfo {year} {2009})}\BibitemShut {NoStop}%
\bibitem [{\citenamefont {Aoki}\ \emph
  {et~al.}(2014{\natexlab{b}})\citenamefont {Aoki}, \citenamefont {Tsuji},
  \citenamefont {Eckstein}, \citenamefont {Kollar}, \citenamefont {Oka},\ and\
  \citenamefont {Werner}}]{57a}%
  \BibitemOpen
  \bibfield  {author} {\bibinfo {author} {\bibfnamefont {H.}~\bibnamefont
  {Aoki}}, \bibinfo {author} {\bibfnamefont {N.}~\bibnamefont {Tsuji}},
  \bibinfo {author} {\bibfnamefont {M.}~\bibnamefont {Eckstein}}, \bibinfo
  {author} {\bibfnamefont {M.}~\bibnamefont {Kollar}}, \bibinfo {author}
  {\bibfnamefont {T.}~\bibnamefont {Oka}},\ and\ \bibinfo {author}
  {\bibfnamefont {P.}~\bibnamefont {Werner}},\ }\bibfield  {title} {\bibinfo
  {title} {Nonequilibrium dynamical mean-field theory and its applications},\
  }\href {https://doi.org/10.1103/RevModPhys.86.779} {\bibfield  {journal}
  {\bibinfo  {journal} {Rev. Mod. Phys.}\ }\textbf {\bibinfo {volume} {86}},\
  \bibinfo {pages} {779} (\bibinfo {year} {2014}{\natexlab{b}})}\BibitemShut
  {NoStop}%
\bibitem [{\citenamefont {Zhu}\ \emph {et~al.}(2013)\citenamefont {Zhu},
  \citenamefont {Liu},\ and\ \citenamefont {Guo}}]{66a}%
  \BibitemOpen
  \bibfield  {author} {\bibinfo {author} {\bibfnamefont {Y.}~\bibnamefont
  {Zhu}}, \bibinfo {author} {\bibfnamefont {L.}~\bibnamefont {Liu}},\ and\
  \bibinfo {author} {\bibfnamefont {H.}~\bibnamefont {Guo}},\ }\bibfield
  {title} {\bibinfo {title} {Quantum transport theory with nonequilibrium
  coherent potentials},\ }\href {https://doi.org/10.1103/PhysRevB.88.205415}
  {\bibfield  {journal} {\bibinfo  {journal} {Phys. Rev. B}\ }\textbf {\bibinfo
  {volume} {88}},\ \bibinfo {pages} {205415} (\bibinfo {year}
  {2013})}\BibitemShut {NoStop}%
\bibitem [{\citenamefont {Kalitsov}\ \emph {et~al.}(2012)\citenamefont
  {Kalitsov}, \citenamefont {Chshiev},\ and\ \citenamefont {Velev}}]{67a}%
  \BibitemOpen
  \bibfield  {author} {\bibinfo {author} {\bibfnamefont {A.~V.}\ \bibnamefont
  {Kalitsov}}, \bibinfo {author} {\bibfnamefont {M.~G.}\ \bibnamefont
  {Chshiev}},\ and\ \bibinfo {author} {\bibfnamefont {J.~P.}\ \bibnamefont
  {Velev}},\ }\bibfield  {title} {\bibinfo {title} {Nonequilibrium coherent
  potential approximation for electron transport},\ }\href
  {https://doi.org/10.1103/PhysRevB.85.235111} {\bibfield  {journal} {\bibinfo
  {journal} {Phys. Rev. B}\ }\textbf {\bibinfo {volume} {85}},\ \bibinfo
  {pages} {235111} (\bibinfo {year} {2012})}\BibitemShut {NoStop}%
\bibitem [{\citenamefont {Dohner}\ \emph {et~al.}(2022)\citenamefont {Dohner},
  \citenamefont {Terletska}, \citenamefont {Tam}, \citenamefont {Moreno},\ and\
  \citenamefont {Fotso}}]{DMFT_CPA_NonEq2022}%
  \BibitemOpen
  \bibfield  {author} {\bibinfo {author} {\bibfnamefont {E.}~\bibnamefont
  {Dohner}}, \bibinfo {author} {\bibfnamefont {H.}~\bibnamefont {Terletska}},
  \bibinfo {author} {\bibfnamefont {K.-M.}\ \bibnamefont {Tam}}, \bibinfo
  {author} {\bibfnamefont {J.}~\bibnamefont {Moreno}},\ and\ \bibinfo {author}
  {\bibfnamefont {H.~F.}\ \bibnamefont {Fotso}},\ }\bibfield  {title} {\bibinfo
  {title} {Nonequilibrium $\text{DMFT}+\mathrm{CPA}$ for correlated disordered
  systems},\ }\href {https://doi.org/10.1103/PhysRevB.106.195156} {\bibfield
  {journal} {\bibinfo  {journal} {Phys. Rev. B}\ }\textbf {\bibinfo {volume}
  {106}},\ \bibinfo {pages} {195156} (\bibinfo {year} {2022})}\BibitemShut
  {NoStop}%
\bibitem [{\citenamefont {Dohner}\ \emph {et~al.}(2023)\citenamefont {Dohner},
  \citenamefont {Terletska},\ and\ \citenamefont
  {Fotso}}]{dohner2023thermalization}%
  \BibitemOpen
  \bibfield  {author} {\bibinfo {author} {\bibfnamefont {E.}~\bibnamefont
  {Dohner}}, \bibinfo {author} {\bibfnamefont {H.}~\bibnamefont {Terletska}},\
  and\ \bibinfo {author} {\bibfnamefont {H.~F.}\ \bibnamefont {Fotso}},\
  }\bibfield  {title} {\bibinfo {title} {Thermalization of a disordered
  interacting system under an interaction quench},\ }\href
  {https://doi.org/10.1103/PhysRevB.108.144202} {\bibfield  {journal} {\bibinfo
   {journal} {Phys. Rev. B}\ }\textbf {\bibinfo {volume} {108}},\ \bibinfo
  {pages} {144202} (\bibinfo {year} {2023})}\BibitemShut {NoStop}%
\bibitem [{\citenamefont {Yan}\ and\ \citenamefont
  {Werner}(2023)}]{DMFTCPA_PWerner}%
  \BibitemOpen
  \bibfield  {author} {\bibinfo {author} {\bibfnamefont {J.}~\bibnamefont
  {Yan}}\ and\ \bibinfo {author} {\bibfnamefont {P.}~\bibnamefont {Werner}},\
  }\bibfield  {title} {\bibinfo {title} {Dynamical mean-field approach to
  disordered interacting systems and applications to the quantum transport
  problem},\ }\href {https://doi.org/10.1103/PhysRevB.108.125143} {\bibfield
  {journal} {\bibinfo  {journal} {Phys. Rev. B}\ }\textbf {\bibinfo {volume}
  {108}},\ \bibinfo {pages} {125143} (\bibinfo {year} {2023})}\BibitemShut
  {NoStop}%
\bibitem [{\citenamefont {Mazzocchi}\ \emph {et~al.}(2024)\citenamefont
  {Mazzocchi}, \citenamefont {Werner},\ and\ \citenamefont
  {Arrigoni}}]{DMFTCPA_Arrigoni}%
  \BibitemOpen
  \bibfield  {author} {\bibinfo {author} {\bibfnamefont {T.~M.}\ \bibnamefont
  {Mazzocchi}}, \bibinfo {author} {\bibfnamefont {D.}~\bibnamefont {Werner}},\
  and\ \bibinfo {author} {\bibfnamefont {E.}~\bibnamefont {Arrigoni}},\
  }\bibfield  {title} {\bibinfo {title} {Impact of disorder and phonons on the
  hubbard bands of mott insulators in strong electric fields},\ }\href
  {https://doi.org/10.1103/PhysRevB.109.045119} {\bibfield  {journal} {\bibinfo
   {journal} {Phys. Rev. B}\ }\textbf {\bibinfo {volume} {109}},\ \bibinfo
  {pages} {045119} (\bibinfo {year} {2024})}\BibitemShut {NoStop}%
\bibitem [{\citenamefont {Poli}\ \emph {et~al.}(2025)\citenamefont {Poli},
  \citenamefont {Fratini}, \citenamefont {Coulter}, \citenamefont {Millis},\
  and\ \citenamefont {Ciuchi}}]{ArinnaCiuchi2025}%
  \BibitemOpen
  \bibfield  {author} {\bibinfo {author} {\bibfnamefont {A.}~\bibnamefont
  {Poli}}, \bibinfo {author} {\bibfnamefont {S.}~\bibnamefont {Fratini}},
  \bibinfo {author} {\bibfnamefont {J.}~\bibnamefont {Coulter}}, \bibinfo
  {author} {\bibfnamefont {A.~J.}\ \bibnamefont {Millis}},\ and\ \bibinfo
  {author} {\bibfnamefont {S.}~\bibnamefont {Ciuchi}},\ }\href
  {https://arxiv.org/abs/2502.17250} {\bibinfo {title} {Stabilization of
  fermi-liquid behavior by interactions in disordered metals}} (\bibinfo {year}
  {2025}),\ \Eprint {https://arxiv.org/abs/2502.17250} {arXiv:2502.17250
  [cond-mat.str-el]} \BibitemShut {NoStop}%
\bibitem [{\citenamefont {Keldysh}(1964)}]{62a}%
  \BibitemOpen
  \bibfield  {author} {\bibinfo {author} {\bibfnamefont {L.~V.}\ \bibnamefont
  {Keldysh}},\ }\bibfield  {title} {\bibinfo {title} {{Diagram technique for
  nonequilibrium processes}},\ }\href@noop {} {\bibfield  {journal} {\bibinfo
  {journal} {Zh. Eksp. Teor. Fiz.}\ }\textbf {\bibinfo {volume} {47}},\
  \bibinfo {pages} {1515} (\bibinfo {year} {1964})}\BibitemShut {NoStop}%
\bibitem [{\citenamefont {Stefanucci}\ and\ \citenamefont {van
  Leeuwen}(2013)}]{63a}%
  \BibitemOpen
  \bibfield  {author} {\bibinfo {author} {\bibfnamefont {G.}~\bibnamefont
  {Stefanucci}}\ and\ \bibinfo {author} {\bibfnamefont {R.}~\bibnamefont {van
  Leeuwen}},\ }\href@noop {} {\emph {\bibinfo {title} {Nonequilibrium Many-Body
  Theory of Quantum Systems: A Modern Introduction}}}\ (\bibinfo  {publisher}
  {Cambridge University Press},\ \bibinfo {year} {2013})\BibitemShut {NoStop}%
\bibitem [{\citenamefont {Rammer}(2007)}]{64a}%
  \BibitemOpen
  \bibfield  {author} {\bibinfo {author} {\bibfnamefont {J.}~\bibnamefont
  {Rammer}},\ }\href {https://doi.org/10.1017/CBO9780511618956} {\emph
  {\bibinfo {title} {Quantum Field Theory of Non-equilibrium States}}}\
  (\bibinfo  {publisher} {Cambridge University Press},\ \bibinfo {year}
  {2007})\BibitemShut {NoStop}%
\bibitem [{\citenamefont {Kadanoff}\ and\ \citenamefont {Baym}(1962)}]{65a}%
  \BibitemOpen
  \bibfield  {author} {\bibinfo {author} {\bibfnamefont {L.}~\bibnamefont
  {Kadanoff}}\ and\ \bibinfo {author} {\bibfnamefont {G.}~\bibnamefont
  {Baym}},\ }\href@noop {} {\emph {\bibinfo {title} {Quantum Statistical
  Mechanics: Green’s Function Methods in Equilibrium and Nonequilibrium
  Problems (1st ed.)}}}\ (\bibinfo  {publisher} {CRC Press},\ \bibinfo {year}
  {1962})\BibitemShut {NoStop}%
\end{thebibliography}%

\end{document}